\documentclass[letterpaper, conference]{IEEEtran}
\IEEEoverridecommandlockouts
\usepackage[left=1.91cm,right=1.91cm,top=2.54cm,bottom=2cm]{geometry}
\usepackage{cite}
\usepackage{amsmath,amssymb,amsfonts}
\usepackage{algorithmic}
\usepackage{graphicx}
\usepackage{textcomp}
\usepackage{xcolor}
\def\BibTeX{{\rm B\kern-.05em{\sc i\kern-.025em b}\kern-.08em
    T\kern-.1667em\lower.7ex\hbox{E}\kern-.125emX}}

\begin{document}

\title{Extended Adaptive Observer for Linear Systems with Overparametrization\\
\thanks{Financial support is in part provided by the Grants Council of the President of the Russian Federation (MD-1787.2022.4).}
}

\author{\IEEEauthorblockN{1\textsuperscript{st} Anton Glushchenko}
\IEEEauthorblockA{\textit{Laboratory No. 7} \\
\textit{V.A. Trapeznikov Institute of Control Sciences of RAS}\\
Moscow, Russia\\
aiglush@ipu.ru}
\and
\IEEEauthorblockN{2\textsuperscript{nd} Konstantin Lastochkin}
\IEEEauthorblockA{\textit{Laboratory No. 7} \\
\textit{V.A. Trapeznikov Institute of Control Sciences of RAS}\\
Moscow, Russia\\
lastconst@ipu.ru}
}

\maketitle

\begin{abstract}
Exponentially stable extended adaptive observer is proposed for a class of linear time-invariant systems with unknown parameters and overparameterization. It allows one to reconstruct unmeasured states and bounded external disturbance produced by a known linear exosystem with unknown initial conditions if a weak requirement of regressor finite excitation is met. In contrast to the existing solutions, the proposed observer reconstructs the original (physical) states of the system rather than the virtual one of its observer canonical form. Simulation results to validate the developed theory are presented.
\end{abstract}


\section{Introduction}
In the 1970s several different approaches to design adaptive observers were proposed in the control literature \cite{b1, b2, b3, b4, b5, b6}. \emph{Carroll} and \emph{Lindorff} \cite{b1} were the first who developed a method of adaptive observation of both unmeasured state and unknown parameters simultaneously. \emph{Luders} and \emph{Narendra} suggested an alternative observer \cite{b2} and later modified it to have a simpler structure \cite{b3}. The paper by \emph{Kudva} and \emph{Narendra} \cite{b4} proposed yet another method, and in \cite{b3, b5} \emph{Narendra} and \emph{Kudva} showed that all these results \cite{b1, b2, b4} could be derived in a unified manner. \emph{Kreisselmeier} \cite{b6} proposed a parametrization that, unlike \cite{b1, b2, b4}, allows one to: \emph{i}) transform the adaptive observer design problem into the one of estimation of a linear regression equation (LRE) unknown parameters; \emph{ii}) separate completely the observer dynamics from the adaptive loop to make it possible to apply any parameter adaptation law.

Adaptive observers developed in \cite{b2, b3, b4, b5, b6} are capable of unmeasured state reconstruction in case the system model is represented in the observer canonical form. The obtained states  $\xi \left( t \right)$ are useless to solve a number of control problems because they are virtual and related to the plant original states \linebreak $x\left( t \right) \in {\mathbb{R}^n}$ through the \emph{unknown} linear transformation $x\left( t \right) = T\xi \left( t \right)$. In contrast to \cite{b1, b3, b4, b5, b6}, in addition to the estimates $\hat \xi \left( t \right)$, for one class of linear systems the observer from \cite{b2} also forms an estimate $\hat T\left( t \right)$ and recovers the original state $x\left( t \right)$ with the help of equation $\hat x\left( t \right) = \hat T\left( t \right)\hat \xi \left( t \right)$. However, the estimate $\hat T\left( t \right)$ can become discontinuous, and the transient quality of the observation error $\tilde x\left( t \right)$ can be arbitrarily bad (see Section VIII of \cite{b1}). In the later papers \cite{b7, b8} and many others devoted to the development of methods to design adaptive observers (and even in the seminal books on adaptive observers for linear systems \cite{b9, b10}), to the best of the authors’ knowledge, the problem of original state $x\left( t \right)$ reconstruction and a linear similarity matrix $T$ identification by adaptive observers has not been touched upon anymore.

In the recent study \cite{b11} a new method to reconstruct the original state vector of linear systems with unknown parameters and overparameterization has been proposed. In this paper, on the basis of the results from \cite{b11}, a new extended adaptive observer is proposed, which additionally allows one to estimate the external disturbance affecting the system.

The following definitions of the heterogeneous mapping and the regressor finite excitation condition are used throughout this study.

{\it \bf Definition 1.} \emph{A mapping ${\cal F}\left( x \right){\rm{:\;}}{\mathbb{R}^{{n_x}}} \to {\mathbb{R}^{{n_{\cal F}} \times {m_{\cal F}}}}$ is heterogeneous of degree ${\ell _{\cal F}} \ge 1$ if there exists ${\Xi _{\cal F}}\left( {\omega \left( t \right)} \right)=\linebreak = {\overline \Xi _{\cal F}}\left( {\omega \left( t \right)} \right)\omega \left( t \right) \in {\mathbb{R}^{{\Delta _{\cal F}} \times {n_x}}}{\rm{,\;}}{\Pi _{\cal F}}\left( {\omega \left( t \right)} \right) \in {\mathbb{R}^{{n_{\cal F}} \times {n_{\cal F}}}}$, and a mapping ${{\cal T}_{\cal F}}\left( {{\Xi _{\cal F}}\left( {\omega \left( t \right)} \right)x} \right){\rm{:\;}}{\mathbb{R}^{{\Delta _{\cal F}}}} \to {\mathbb{R}^{{n_{\cal F}} \times {m_{\cal F}}}}$ such that for all $\left| {\omega \left( t \right)} \right| > 0$ and $x \in {\mathbb{R}^{{n_x}}}$ the following conditions hold:}
\begin{equation}\label{eq1}
\begin{array}{c}
{\Pi _{\cal F}}\left( {\omega \left( t \right)} \right){\cal F}\left( x \right) = {{\cal T}_{\cal F}}\left( {{\Xi _{\cal F}}\left( {\omega \left( t \right)} \right)x} \right){\rm{, }}\\
{\rm{det}}\left\{ {{\Pi _{\cal F}}\left( {\omega \left( t \right)} \right)} \right\} \ge {\omega ^{{\ell _{_{\cal F}}}}}\left( t \right)\!{\rm{,}}{\Xi _{\cal F}}_{ij}\!\left( {\omega \left( t \right)} \right) = {c_{ij}}{\omega ^\ell }\left( t \right)\!{\rm{,}}\\
{c_{ij}} \in \left\{ {0,{\rm{ 1}}} \right\}{\rm{,\;}}\ell  > 0.
\end{array}
\end{equation}

For instance, the mapping ${\cal F}\left( x \right) = {\rm{col}}\left\{ {{x_1}{x_2}{\rm{,\;}}{x_1}} \right\}$ with ${\Pi _{\cal F}}\left( \omega  \right) = {\rm{diag}}\left\{ {{\omega ^2}{\rm{,\;}}\omega } \right\}{\rm{,\;}}{\Xi _{\cal F}}\left( \omega  \right) = {\rm{diag}}\left\{ {\omega {\rm{,\;}}\omega } \right\}$ is heterogeneous of degree ${\ell _{\cal F}} = 3.$

{\it \bf Definition 2.} \emph{The regressor $\varphi \left( t \right) \in {\mathbb{R}^n}$ is finitely exciting $\varphi \left( t \right) \in {\rm{FE}}$  over the time range $\left[ {t_r^ + {\rm{;\;}}{t_e}} \right]$  if there exists $t_r^ +  \ge 0$, ${t_e} > t_r^ +$ and $\alpha$ such that the following inequality holds:}
\begin{equation}\label{eq2}
\int\limits_{t_r^ + }^{{t_e}} {\varphi \left( \tau  \right){\varphi ^{\rm{T}}}\left( \tau  \right)d} \tau  \ge \alpha {I_n}{\rm{,}}
\end{equation}
\emph{where $\alpha > 0$ is the excitation level, $I_{n}$ is an identity matrix.}

\newgeometry{left=1.91cm,right=1.91cm,top=1.91cm,bottom=2cm}

\section{Problem Statement}

A class of linear time-invariant systems with overparametrization and external bounded disturbance is considered:
\begin{equation}\label{eq3}
\begin{array}{l}
\dot x\left( t \right) = A\left( \theta  \right)x\left( t \right) + B\left( \theta  \right)u\left( t \right) + D\left( \theta  \right)\delta \left( t \right){\rm{,}}\\
y\left( t \right) = {C^{\rm{T}}}x\left( t \right){\rm{,\;}}x\left( {{t_0}} \right) = {x_0}{\rm{,}}
\end{array}
\end{equation}
where $x\left( t \right) \in {\mathbb{R}^n}$ is a vector of original (physical) state of the system with unknown initial conditions ${x_0}$, $\delta \left( t \right) \in \mathbb{R}$ stands for a bounded external perturbation, $C \in {\mathbb{R}^n}$ is a known vector, $A{\rm{:\;}}{\mathbb{R}^{{n_\theta }}} \to {\mathbb{R}^{n \times n}}{\rm{,\;}}B{\rm{:\;}}{\mathbb{R}^{{n_\theta }}} \to {\mathbb{R}^n}{\rm{,\;}}D{\rm{:\;}}{\mathbb{R}^{{n_\theta }}} \to {\mathbb{R}^n}$ denote known mappings with unknown original (physical) parameters $\theta  \in {\mathbb{R}^{{n_\theta }}}$. Only control $u\left( t \right) \in \mathbb{R}$ and output $y\left( t \right) \in {\mathbb{R}}$ signals are measurable.

The following assumptions related to the control signal and disturbance are adopted.

{\it \bf Assumption 1.} \emph{The control signal $u\left( t \right)$ is chosen so that for $t \ge {t_0}$ it ensures existence and boundedness of all trajectories of the system \eqref{eq3}.}

{\it \bf Assumption 2.} \emph{The disturbance $\delta \left( t \right)$ is a bounded and continuous signal produced by a linear time-invariant exosystem:}
\begin{equation}\label{eq4}
\begin{array}{l}
{{\dot x}_\delta }\left( t \right) = {{\cal A}_\delta }{x_\delta }\left( t \right){\rm{,\;}}{x_\delta }\left( {{t_0}} \right) = {x_{\delta 0}}{\rm{,}}\\
\delta \left( t \right) = h_\delta ^{\rm{T}}{x_\delta }\left( t \right){\rm{,}}
\end{array}
\end{equation}
\emph{where ${x_\delta }\left( t \right) \in {R^{{n_\delta }}}$ is the exosystem state vector with unknown initial conditions ${x_{\delta 0}}\left( {{t_0}} \right)$, ${h_\delta } \in {R^{{n_\delta }}}{\rm{,\;}}{{\cal A}_\delta } \in {R^{{n_\delta } \times {n_\delta }}}$ are known vector and matrix such that the pair $\left( {h_\delta ^{\rm{T}}{\rm{,\;}}{{\cal A}_\delta }} \right)$ is observable.}

Under Assumption 2, the model \eqref{eq3} is extended with \eqref{eq4}:
\begin{equation}\label{eq5}
\begin{array}{r}
{{\dot x}_e}\left( t \right) = \left( {{A_e}\left( \theta  \right) + {A_\delta }} \right){x_e}\left( t \right) + {B_e}\left( \theta  \right)u\left( t \right) = \\
 = {\Phi ^{\rm{T}}}\left( {{x_e}{\rm{, }}u} \right){\Theta _{AB}}\left( \theta  \right) + {A_\delta }{x_e}\left( t \right){\rm{,}}\\
y\left( t \right) = C_e^{\rm{T}}{x_e}\left( t \right){\rm{,\;}}{x_e}\left( {{t_0}} \right) = { {\begin{bmatrix}
{{x_0}}&{{x_{\delta 0}}}
\end{bmatrix}}^{\rm{T}}}{\rm{,}}
\end{array}
\end{equation}
where
\begin{gather*}
{A_e}\left( \theta  \right) = {\begin{bmatrix}
{A\left( \theta  \right)}&{D\left( \theta  \right)h_\delta ^{\rm{T}}}\\
{{0_{{n_\delta } \times n}}}&{{0_{{n_\delta } \times {n_\delta }}}}
\end{bmatrix}} {\rm{,\;}}{A_\delta } = {\begin{bmatrix}
{{0_{n \times n}}}&{{0_{n \times {n_\delta }}}}\\
{{0_{{n_\delta } \times n}}}&{{{\cal A}_\delta }}
\end{bmatrix}}{\rm{, }}\\
{B_e}\left( \theta  \right) = {\begin{bmatrix}
{B\left( \theta  \right)}\\
{{0_{{n_\delta }}}}
\end{bmatrix}} {\rm{,\;}}{x_e}\left( t \right) = {\begin{bmatrix}
{x\left( t \right)}\\
{{x_\delta }\left( t \right)}
\end{bmatrix}} {\rm{,\;}}{C_e} = {\begin{bmatrix}
C\\
{{0_{{n_\delta }}}}
\end{bmatrix}}{\rm{,}}\\
{\Phi ^{\rm{T}}}\left( {{x_e}{\rm{,\;}}u} \right) = {\begin{bmatrix}
{{I_{n_{e}}} \otimes x_e^{\rm{T}}\left( t \right)}&{{I_{n_{e}}} \otimes {u^{\rm{T}}}\left( t \right)}
\end{bmatrix}}{{\cal D}_\Phi } \in {\mathbb{R}^{{n_{e}} \times {n_\Theta }}}{\rm{, }}\\
{\Theta _{AB}}\left( \theta  \right) = {{\cal L}_\Phi }{ {\begin{bmatrix}
{ve{c^{\rm{T}}}\left( {A_e^{\rm{T}}\left( \theta  \right)} \right)}&{B_e^{\rm{T}}\left( \theta  \right)}
\end{bmatrix}}^{\rm{T}}} \in {\mathbb{R}^{{n_\Theta }}}{\rm{,}}
\end{gather*}
${\Theta _{AB}}{\rm{:\;}} {\mathbb{R}^{{n_\theta }}} \!\to\! {\mathbb{R}^{{n_\Theta }}}$ is a known mapping s.t. $n \!\ge\! {n_\Theta } \!\ge\! {n_\theta }$, \linebreak${{\cal D}_\Phi } \in {\mathbb{R}^{ {n_{e}}\left( {n_{e} + 1} \right) \times {n_\Theta }}}{\rm{,\;\;}}{{\cal L}_\Phi } \in {\mathbb{R}^{{n_\Theta } \times {n_{e}}\left( {n_{e} + 1} \right)}}$ with \linebreak  $n_{e}=n+n_{\delta}$ stand for known duplication and elimination matrices, respectively\footnote{Dependencies from $\theta$ and $t$ can be further suppressed for the sake of brevity.}.

Owing to the duality of observation and control problems, the following generalized pole placement control theory assumption \cite{b12, b13} is also adopted.

{\it \bf Assumption 3.} \emph{The pair $\left( {A_e^{\rm{T}}\left( \theta  \right) + A_\delta ^{\rm{T}}{\rm{,\;}}{C_e}} \right)$ is controllable, and there exists a known state matrix $\Gamma  \in {\mathbb{R}^{n_{e} \times n_{e}}}$ of an exosystem:}
\begin{equation}\label{eq6}
\begin{array}{l}
\dot \varsigma \left( t \right) = \Gamma \varsigma \left( t \right){\rm{,}}\\
v\left( t \right) = B_e^{\rm{T}}\left( \theta  \right)\varsigma \left( t \right){\rm{,}}
\end{array}
\end{equation}
\emph{such that $\sigma \left\{ {A_e^{\rm{T}}\left( \theta  \right) + A_\delta ^{\rm{T}}} \right\} \cap \sigma \left\{ \Gamma  \right\} = 0$ and the pair $\left( {B_e^{\rm{T}}\left( \theta  \right){\rm{,\;}}\Gamma } \right)$ is observable.}

According to \cite{b12, b13}, under Assumption 3  there exists a vector $L\left( \theta  \right) \in {\mathbb{R}^{n_{e}}}$, which makes the algebraic spectrum $\sigma \{.\}$ of the matrix $A_e^{\rm{T}}\left( \theta  \right)+A_\delta ^{{\rm{T}}} - {C_e}{L^{\rm{T}}}\left( \theta  \right)$ be equal to the desired one. So, in accordance with Assumptions 1-3 an extended adaptive observer is introduced as follows:
\begin{equation}\label{eq7}
\begin{gathered}
{{\dot {\hat {x}}}_e}\left( t \right) = {\Phi ^{\rm{T}}}\left( {{{\hat x}_e}{\rm{,\;}}u} \right){{\hat \Theta }_{AB}}\left( t \right) + {A_\delta }{{\hat x}_e}\left( t \right) - \\
{\rm{\;\;\;\;\;\;\;\;\;\;\;\;\;\;\;\;\;\;\;\;\;\;}} - \hat L\left( t \right)\left( {\hat y\left( t \right) - y\left( t \right)} \right).
\end{gathered}
\end{equation}

{\bf{Goal.}} The observer \eqref{eq7} is required to be augmented with the estimation laws, which ensure that the following equalities hold:
\begin{equation}\label{eq8}
\begin{array}{c}
\mathop {{\rm{lim}}}\limits_{t \to \infty } \left\| {{{\tilde x}_e}\left( t \right)} \right\| = 0{\rm{ }}\left( {exp } \right){\rm{,\;}}\mathop {{\rm{lim}}}\limits_{t \to \infty } \left\| {\tilde \kappa \left( t \right)} \right\| = 0{\rm{\;}}\left( {exp } \right){\rm{,}}\\
\tilde \kappa \left( t \right) = {{\begin{bmatrix}
{\tilde \Theta _{AB}^{\rm{T}}\left( t \right)}&{{{\tilde L}^{\rm{T}}}\left( t \right)}
\end{bmatrix}}^{\rm{T}}}{\rm{,}}
\end{array}
\end{equation}
where ${\tilde x_e}\left( t \right) = {\hat x_e}\left( t \right) - {x_e}\left( t \right)$ stands for the state observation error for the system \eqref{eq5}, ${\tilde \Theta _{AB}}\left( t \right) = {\hat \Theta _{AB}}\left( t \right) - {\Theta _{AB}}\left( \theta  \right)$ is the parameter estimation error, $\tilde L\left( t \right) = \hat L\left( t \right) - L\left( \theta  \right)$ denotes the estimation error of $L\left( \theta  \right)$.

\section{Main Result}

The proposed solution to the problem \eqref{eq8} consists of a parametrization scheme of the LRE with respect to the unknown parameters $\kappa$ and the identification procedure to obtain the estimates $\hat \kappa \left( t \right)$ required to implement \eqref{eq7}.

\subsection{Parametrization}

To obtain the LRE with respect to $\kappa$, first of all, using the nonsingular transformation $\xi \left( t \right) = T\left( \theta  \right)x\left( t \right)$, the system \eqref{eq3} is represented in the observer canonical form\footnote{According to \cite{b1}, each completely observable system can be transformed into \eqref{eq9}.}:
\begin{equation}\label{eq9}
\dot \xi \left( t \right)\! =\! {A_0}\xi \left( t \right) + {\psi _a}\left( \theta  \right)y\left( t \right) + {\psi _b}\left( \theta  \right)u\left( t \right) + {\psi _d}\left( \theta  \right)\delta \left( t \right){\rm{,}}
\end{equation}
\begin{equation}\label{eq10}
y\left( t \right) = {C^{\rm{T}}}x\left( t \right) = C_0^{\rm{T}}\xi \left( t \right){\rm{,\;}}\xi \left( {{t_0}} \right) = {\xi _0}{\rm{,}}
\end{equation}
where
\begin{gather*}
    \begin{array}{c}
{\psi _a}\left( \theta  \right) = T\left( \theta  \right)A\left( \theta  \right){T^{ - 1}}\left( \theta  \right){C_0}{\rm{,\;}}{\psi _b}\left( \theta  \right) = T\left( \theta  \right)B\left( \theta  \right){\rm{,}}\\
{\psi _d}\left( \theta  \right) = T\left( \theta  \right)D\left( \theta  \right){\rm{,\;}}C_0^{\rm{T}} = { C} ^{{\rm{T}}}{T^{ - 1}}\left( \theta  \right){\rm{,}}\\
{T^{ - 1}}\left( \theta  \right) = {\begin{bmatrix}
{{A^{n - 1}}\left( \theta  \right){{\cal O}_n}}&{{A^{n - 2}}\left( \theta  \right){{\cal O}_n}}& \cdots &{{{\cal O}_n}}
\end{bmatrix}} {\rm{,}}
\end{array}
\end{gather*}
${{\cal O}_n}\left( \theta  \right)$ is the \emph{n}$^{\rm th}$ column of the matrix that is an inverse one to ${{\cal O}^{ - 1}}\left( \theta  \right) = {{\begin{bmatrix}
  C&{{A^{\text{T}}\left( \theta  \right)}C}& \cdots &{{{\left( {{A^{n - 1}\left( \theta  \right)}} \right)}^{\text{T}}}C} 
\end{bmatrix}}^{\text{T}}}{\rm{,}}$ $\xi \left( t \right) \in {\mathbb{R}^n}$ are states of the system represented in the observer canonical form with unknown initial conditions ${\xi _0}$, the vector ${C_0} \in {\mathbb{R}^n}$ and mappings ${\psi _a}{\rm{,\;}}{\psi _b}{\rm{,\;}}{\psi _d}{\rm{:\;}}{\mathbb{R}^{{n_\theta }}} \to {\mathbb{R}^n}$ are known.

The following parametrization can be obtained for the unknown parameters $\eta \left( \theta  \right) = {\rm{col}}\left\{ {{\psi _a}\left( \theta  \right){\rm{,\;}}{\psi _b}\left( \theta  \right)} \right\}$ of equation \eqref{eq9} in case Assumption 2 holds.

{\bf{Lemma 1.}} \emph{Let ${t_\epsilon} > {t_0}$ be a sufficiently large predefined time instance, then for all $t \ge {t_{\epsilon}}$  the unknown parameters $\eta \left( \theta  \right)$ satisfy the LRE:}
\begin{equation}\label{eq11}
\begin{array}{c}
{\cal Y}\left( t \right) = \Delta \left( t \right)\eta \left( \theta  \right),\\
\end{array}
\end{equation}
\begin{gather*}
    {\cal Y}\left( t \right) = k\left( t \right) \cdot {\rm{adj}}\left\{ {\varphi \left( t \right)} \right\}q\left( t \right){\rm{,\;}}\Delta \left( t \right) = k\left( t \right) \cdot {\rm{det}}\left\{ {\varphi \left( t \right)} \right\}{\rm{,}}
\end{gather*}
\emph{where}
\begin{gather*}
\begin{array}{l}
\dot q\left( t \right) = \int\limits_{t_{\epsilon}}^t {{e^{ - k_{2} \tau }}{{\overline \varphi }_f}\left( \tau  \right)\left( \begin{array}{c}
\overline q\left( \tau  \right) - {k_1}{{\overline q}_f}\left( \tau  \right) - \\
 - {\beta ^{\rm{T}}}\left( {{F_f}\left( \tau  \right) + l{y_f}\left( \tau  \right)} \right)
\end{array} \right)d\tau },
\end{array}
\end{gather*}
\begin{equation}\label{eq12}
\begin{array}{l}
q\left( t_{\epsilon} \right) = {0_{{\rm{2}}n}},\\
\dot \varphi \left( t \right) = \int\limits_{t_{\epsilon}}^t {{e^{ - k_{2} \tau }}{{\overline \varphi }_f}\left( \tau  \right)\overline \varphi _f^{\rm{T}}\left( \tau  \right)d\tau } {\rm{,\;}}\varphi \left( t_{\epsilon}  \right) = {0_{{\rm{2}}n \times {\rm{2}}n}},
\end{array} 
\end{equation}
\begin{equation}\label{eq13}
\begin{array}{l}
{{\dot {\overline q}}_f}\left( t \right) =  - {k_1}{{\overline q}_f}\left( t \right) + \overline q\left( t \right){\rm{,\;}}{{\overline q}_f}\left( {{t_0}} \right) = 0,\\
{{\dot {\overline \varphi} }_f}\left( t \right) =  - {k_1}{{\overline \varphi }_f}\left( t \right) + \overline \varphi \left( t \right){\rm{,\;}}{{\overline \varphi }_f}\left( {{t_0}} \right) = {0_{{\rm{2}}n}},\\
{{\dot F}_f}\left( t \right) =  - {k_1}{F_f}\left( t \right) + F\left( t \right){\rm{,\;}}{F_f}\left( {{t_0}} \right) = {0_{{n_\delta }}},\\
{{\dot y}_f}\left( t \right) =  - {k_1}{y_f}\left( t \right) + y\left( t \right){\rm{,\;}}{y_f}\left( {{t_0}} \right) = 0,
\end{array}
\end{equation}
\begin{equation}\label{eq14}
\begin{array}{c}
\overline q\left( t \right) = y\left( t \right) - C_0^{\rm{T}}z{\rm{,\;}}\overline \varphi \left( t \right) = \left[ {\begin{array}{*{20}{c}}
{{{\dot \Omega }^{\rm{T}}}{C_0} + {N^{\rm{T}}}\beta }\\
{{{\dot P}^{\rm{T}}}{C_0} + {H^{\rm{T}}}\beta }
\end{array}} \right]{\rm{,}}\\
\dot z\left( t \right) = {A_K}z\left( t \right) + Ky\left( t \right){\rm{,\;}}z\left( {{t_0}} \right) = {0_n}{\rm{,}}\\
\dot \Omega \left( t \right) = {A_K}\Omega \left( t \right) + {I_n}y\left( t \right){\rm{,\;}}\Omega \left( {{t_0}} \right) = {0_{n \times n}}{\rm{,}}\\
\dot P\left( t \right) = {A_K}P\left( t \right) + {I_n}u\left( t \right){\rm{,\;}}P\left( {{t_0}} \right) = {0_{n \times n}}{\rm{,}}\\
\dot F\left( t \right) \!=\! GF\left( t \right) + Gly\left( t \right) - lC_0^{\rm{T}}\dot z\left( t \right){\rm{,\;}}F\left( {{t_0}} \right) \!=\! {0_{{n_\delta }}}{\rm{,}}\\
\dot H\left( t \right) = GH\left( t \right) - lC_0^{\rm{T}}\dot P\left( t \right){\rm{,\;}}H\left( {{t_0}} \right) = {0_{{n_\delta } \times n}}{\rm{,}}\\
\dot N\left( t \right) = GN\left( t \right) - lC_0^{\rm{T}}\dot \Omega \left( t \right){\rm{,\;}}N\left( {{t_0}} \right) = {0_{{n_\delta } \times n}}{\rm{,}}
\end{array}
\end{equation}
\emph{and, if $\overline \varphi \left( t \right) \in {\rm{FE}}$ over the time range $\left[ {{t_{\epsilon}}{\rm{;\;}}{t_e}} \right]$, for all $t_{\epsilon} \ge {t_e}$ it holds that ${\Delta _{{\rm{max}}}} \ge \Delta \left( t \right) \ge {\Delta _{{\rm{min}}}}$.}

\emph{Here $k\left( t \right) > 0$ is the amplifier that can be chosen to be time-varying, ${k_1} > 0,{\rm{\;}}{k_2} > 0$ are filters time constants, ${A_K} = {A_0} - KC_0^{\rm{T}}{\rm{,\;}}G$ stand for stable matrices of appropriate dimensions, the vector $l \in {\mathbb{R}^{{n_\delta }}}$ is such that the pair $\left( {G{\rm{,\;}}l} \right)$ is controllable, and $G$ is chosen in accordance with the condition $\sigma \left\{ {{{\cal A}_\delta }} \right\} \cap \sigma \left\{ G \right\} = 0$, the parameter $\beta  \in {R^{{n_\delta }}}$ is a solution of the following equations:}
\begin{gather*}
    \begin{array}{l}
{M_\delta }{{\cal A}_\delta } - G{M_\delta } = l\overline h_\delta ^{\rm{T}}{\rm{, }}\overline h_\delta ^{\rm{T}} = h_\delta ^{\rm{T}}{{\cal A}_\delta }{\rm{,}}\\
\beta  = \overline h_\delta ^{\rm{T}}M_\delta ^{ - 1}.
\end{array}
\end{gather*}

{\it Proof of Lemma 1 is postponed to Appendix.}

Using the measurable signals $u\left( t \right){\rm{,\;}}y\left( t \right)$ and parametrization \eqref{eq11}, if $\overline \varphi \left( t \right) \in {\rm{FE}}$, then only parameters ${\psi _a}{\rm{,\;}}{\psi _b}$ of the characteristic polynomial of the transfer function ${W_{uy}}\left( s \right) =\linebreak = {C^{\rm{T}}}{\left( {s{I_n} - A\left( \theta  \right)} \right)^{ - 1}}B\left( \theta  \right)$ can be identified \cite{b1, b2, b3, b4, b5, b6, b9, b10}. However, it should be taken into consideration that, following the problem statement, the parameters ${\Theta _{AB}}{\rm{,\;}}L$ depend on the physical parameters $\theta$ in nonlinear and known manner. In their turn, the parameters ${\psi _a}{\rm{,\;}}{\psi _b}$ of the transfer function ${W_{uy}}\left( s \right)$ characteristic polynomial also nonlinearly depends on $\theta$. Then, if the following condition is met:
\begin{equation}\label{eq15}
\begin{array}{c}
{{\rm{det}} ^2}\left\{ {{\nabla _\theta }{\psi _{ab}}\left( \theta  \right)} \right\} > 0,\\{\psi _{ab}}\left( \theta  \right) = {{\cal L}_{ab}}\eta \left( \theta  \right){\rm{:\;}}{\mathbb{R}^{{n_\theta }}} \to {\mathbb{R}^{{n_\theta }}}{\rm{,}}
\end{array}
\end{equation}
owing to the inverse function theorem, there exists an inverse transform $\theta  = {\cal F}\left( {{\psi _{ab}}} \right){\rm{:\;}}{\mathbb{R}^{{n_\theta }}} \to {\mathbb{R}^{{n_\theta }}}$. Therefore, we can: \linebreak \emph{i}) obtain parameters of the system ${\Theta _{AB}}$ and observer $L$ on the basis of ${\psi _{ab}}$, \emph{ii}) implement the adaptive observer \eqref{eq7}, which forms the estimate ${\hat x_e}\left( t \right)$.

In this study, assuming that condition \eqref{eq15} is met, in order to make the problem of unmeasured state $x\left( t \right)$ reconstruction solvable we additionally introduce the following hypothesis about the mapping ${\psi _{ab}}\left( \theta  \right)$.

{\bf{Hypothesis 1.}} \emph{There exist heterogeneous in the sense of \eqref{eq1} mappings ${\cal G}\left( {{\psi _{ab}}} \right){\rm{:\;}}{\mathbb{R}^{{n_\theta }}} \to {\mathbb{R}^{{n_\theta } \times {n_\theta }}}$, ${\cal S}\left( {{\psi _{ab}}} \right){\rm{:\;}}{\mathbb{R}^{{n_\theta }}} \to {\mathbb{R}^{{n_\theta }}}$  such that:}
\begin{equation}\label{eq17}
\begin{array}{c}
{\cal S}\left( {{\psi _{ab}}} \right) = {\cal G}\left( {{\psi _{ab}}} \right){\cal F}\left( {{\psi _{ab}}} \right) = {\cal G}\left( {{\psi _{ab}}} \right)\theta {\rm{,}}\\
\end{array}
\end{equation}
\begin{gather*}
    {\Pi _\theta }\left( {\Delta \left( t \right)} \right){\cal G}\left( {{\psi _{ab}}} \right) = {{\cal T}_{\cal G}}\left( {{\Xi _{\cal G}}\left( {\Delta \left( t \right)} \right){\psi _{ab}}} \right){\rm{:\;}}{\mathbb{R}^{{\Delta _{\cal G}}}} \to {\mathbb{R}^{{n_\theta } \times {n_\theta }}}{\rm{,}}\\
{\Pi _\theta }\left( {\Delta \left( t \right)} \right){\cal S}\left( {{\psi _{ab}}} \right) = {{\cal T}_{\cal S}}\left( {{\Xi _{\cal S}}\left( {\Delta \left( t \right)} \right){\psi _{ab}}} \right){\rm{:\;}}{\mathbb{R}^{{\Delta _{\cal S}}}} \to {\mathbb{R}^{{n_\theta }}}{\rm{,}}
\end{gather*}
\emph{where $\det \left\{ {{\Pi _\theta }\left( {\Delta \left( t \right)} \right)} \right\} \ge {\Delta ^{{\ell _\theta }}}\left( t \right){\rm{,\;}}rank\left\{ {{\cal G}\left( {{\psi _{ab}}} \right)} \right\} = \linebreak = {n_\theta }{\rm{,\;}}{\ell _\theta } \ge 1$, ${\Xi _{\cal G}}\left( {\Delta \left( t \right)} \right) \in {R^{{\Delta _{\cal G}} \times {n_\theta }}}$, ${\Xi _{\cal S}}\left( {\Delta \left( t \right)} \right) \in {\mathbb{R}^{{\Delta _{\cal S}} \times {n_\theta }}}$, and all mappings are known.}

If Hypothesis 1 is met, then, owing to the property ${\Xi _{\left( . \right)}}\left( {\Delta \left( t \right)} \right) = {\overline \Xi _{\left( . \right)}}\left( {\Delta \left( t \right)} \right)\Delta \left( t \right)$, equation \eqref{eq11} is transformed into the LRE with respect to $\theta$:
\begin{equation}\label{eq18}
\begin{array}{c}
{{\cal Y}_\theta }\left( t \right) = {{\cal M}_\theta }\left( t \right)\theta ,\\
\end{array}
\end{equation}
\begin{gather*}
    {{\cal Y}_\theta }\left( t \right) := {\rm{adj}}\left\{ {{{\cal T}_{\cal G}}\left( {{{\overline \Xi }_{\cal G}}\left( \Delta  \right){{\cal Y}_{ab}}} \right)} \right\}{{\cal T}_{\cal S}}\left( {{{\overline \Xi }_{\cal S}}\left( \Delta  \right){{\cal Y}_{ab}}} \right){\rm{,}}\\
{{\cal M}_\theta }\left( t \right) := {\rm{det}}\left\{ {{{\cal T}_{\cal G}}\left( {{{\overline \Xi }_{\cal G}}\left( \Delta  \right){{\cal Y}_{ab}}} \right)} \right\}{\rm{,\;}}{{\cal Y}_{ab}}\left( t \right) := {{\cal L}_{ab}}{\cal Y}\left( t \right).
\end{gather*}

Having \eqref{eq18} at hand and assuming the mapping ${\Theta _{AB}}$ to be heterogeneous in the sense of \eqref{eq1}, the equation \eqref{eq18} can be transformed into LRE with respect to ${\Theta _{AB}}\left( \theta  \right)$.

{\bf{Hypothesis 2.}} \emph{The mapping ${\Theta _{AB}}\left( \theta  \right){\rm{:\;}}{\mathbb{R}^{{n_\theta }}} \to {\mathbb{R}^{{n_\Theta }}}$ is heterogeneous in the sense of \eqref{eq1} such that:}
\begin{equation}\label{eq19}
{\Pi _\Theta }\left( {{{\cal M}_\theta }} \right){\Theta _{AB}}\left( \theta  \right) \!=\! {{\cal T}_\Theta }\left( {{\Xi _\Theta }\left( {{{\cal M}_\theta }} \right)\theta } \right){\rm{:\;}}{\mathbb{R}^{{\Delta _\Theta }}} \to {\mathbb{R}^{{n_\Theta }}}{\rm{,}}
\end{equation}
\emph{where ${\rm{det}} \left\{ {{\Pi _\Theta }\left( {{{\cal M}_\theta }\left( t \right)} \right)} \right\} \ge {\cal M}_\theta ^{{\ell _\Theta }}\left( t \right){\rm{,\;}}{\ell _\Theta } \ge 1$, $ {\Xi _\Theta }\left( {{{\cal M}_\theta }\left( t \right)} \right) \in {\mathbb{R}^{{\Delta _\Theta } \times {n_\theta }}}$ and all mappings are known.}

Using \eqref{eq19} and the property ${\Xi _\Theta }\left( {{{\cal M}_\theta }\left( t \right)} \right) = \linebreak = {\overline \Xi _\Theta }\left( {{{\cal M}_\theta }\left( t \right)} \right){{\cal M}_\theta }\left( t \right)$  in a way that is similar to \eqref{eq18}, it is obtained that:
\begin{equation}\label{eq20}
\begin{array}{c}
{{\cal Y}_{AB}}\left( t \right) = {{\cal M}_{AB}}\left( t \right){\Theta _{AB}}\left( \theta  \right),\\
{{\cal Y}_{AB}}\left( t \right){\rm{:}} = {\rm{adj}}\left\{ {{\Pi _\Theta }\left( {{{\cal M}_\theta }} \right)} \right\}{{\cal T}_\Theta }\left( {{{\overline \Xi }_\Theta }\left( {{{\cal M}_\theta }} \right){{\cal Y}_\theta }} \right){\rm{,}}\\
{{\cal M}_{AB}}\left( t \right){\rm{:}} = {\rm{det}}\left\{ {{\Pi _\Theta }\left( {{{\cal M}_\theta }\left( t \right)} \right)} \right\}.
\end{array}
\end{equation}

Considering Assumption 1, the regression equation \eqref{eq20} is then converted into the one with respect to $L\left( \theta  \right)$.

{\bf{Lemma 2.}} \emph{If Hypotheses 1-2 and Assumption 3 are met, then the parameters $L\left( \theta  \right)$ satisfy the following LRE:}
\begin{equation}\label{eq21}
\begin{array}{c}
{{\cal Y}_L}\left( t \right) = {{\cal M}_L}\left( t \right)L\left( \theta  \right),\\
\end{array}
\end{equation}
\begin{gather*}
    {{\cal Y}_L}\left( t \right){\rm{:}} = {\rm{adj}}\left\{ {{{\cal T}_{\cal P}}\left( {{{\overline \Xi }_{\cal P}}\left( {{{\cal M}_{AB}}} \right){{\cal Y}_\vartheta }} \right)} \right\}{{\cal T}_{\cal Q}}\left( {{{\overline \Xi }_{\cal Q}}\left( {{{\cal M}_{AB}}} \right){{\cal Y}_\vartheta }} \right){\rm{,}}\\
{{\cal M}_L}\left( t \right){\rm{:}} = {\rm{det}}\left\{ {{{\cal T}_{\cal P}}\left( {{{\overline \Xi }_{\cal P}}\left( {{{\cal M}_{AB}}} \right){{\cal Y}_\vartheta }\left( t \right)} \right)} \right\}{\rm{,}}\\
{{\cal Y}_\vartheta }\left( t \right){\rm{:}} = {{\begin{bmatrix}
{{{\cal Y}_{AB}}}&{{{\cal M}_{AB}}ve{c^{\rm{T}}}\left( \Gamma  \right)}&{{{\cal M}_{AB}}ve{c^{\rm{T}}}\left( {{A_\delta }} \right)}
\end{bmatrix}}^{\rm{T}}}{\rm{,}}
\end{gather*}
\emph{where}
\begin{equation}\label{eq22}
\begin{array}{c}
{{\cal T}_{\cal P}}\left( {{\Xi _{\cal P}}\left( {{{\cal M}_{AB}}} \right)\vartheta } \right) = \\ = ve{c^{ - 1}}\left\{ {{{\cal M}_{AB}}{\rm{adj}}\left\{ { - {{\cal M}_{AB}}{\Gamma ^{\rm{T}}} \otimes {I_{n_{e}}} + } \right.} \right.\\
 + \left. {{I_{n_{e}}} \otimes \left( {ve{c^{ - 1}}\left\{ {{{\cal L}_{{A^{\rm{T}}}}}{{\cal D}_\Phi }{{\cal Y}_{AB}}} \right\} + {{\cal M}_{AB}}{A_\delta }} \right)} \right\} \times \\
 \times {\left. {vec\left( {C{{\left( {{{\cal L}_B}{{\cal D}_\Phi }{{\cal Y}_{AB}}} \right)}^{\rm{T}}}} \right)} \right\}^{\rm{T}}}{\rm{,}}\\
{{\cal T}_{\cal Q}}\left( {{\Xi _{\cal Q}}\left( {{{\cal M}_{AB}}} \right)\vartheta } \right) = \\
 =\! {\rm{det}}\left\{ {{I_{n_{e}}} \otimes \left( {ve{c^{ - 1}}\left\{ {{{\cal L}_{{A^{\rm{T}}}}}{{\cal D}_\Phi }{{\cal Y}_{AB}}} \right\} \!+\! {{\cal M}_{AB}}{A_\delta }} \right) \!- } \right.\\
 - \left. {{{\cal M}_{AB}}{\Gamma ^{\rm{T}}} \otimes {I_{n_{e}}}} \right\}{{\cal L}_B}{{\cal D}_\Phi }{{\cal Y}_{AB}}.
\end{array}
\end{equation}
\emph{${{\cal L}_{{A^{\rm{T}}}}}{\rm{,\;}}{{\cal L}_B}$ are the matrices to extract the vectors $vec\left( {A_e^{\rm{T}}\left( \theta  \right)} \right){\rm{,\;}}{B_e}\left( \theta  \right)$ from ${\Theta _{AB}}\left( \theta  \right)$.}

\emph{Proof of Lemma 2 is presented in Appendix.}

As a result, the combination of LRE \eqref{eq20} and \eqref{eq21} allows one to obtain the required LRE with respect to $\kappa$:
\begin{equation}\label{eq23}
\begin{array}{c}
{{\cal Y}_\kappa }\left( t \right) = {{\cal M}_\kappa }\left( t \right)\kappa {\rm{,}}\\
\end{array}
\end{equation}
\begin{gather*}
    {{\cal Y}_\kappa }\left( t \right) = {\rm{adj}}\left\{ {{\rm{bdiag}}\left\{ {{{\cal M}_{AB}}{I_{{n_\Theta }}}{\rm{,\;}}{{\cal M}_L}{I_{n_{e}}}} \right\}} \right\}{\begin{bmatrix}
{{{\cal Y}_{AB}}}\\
{{{\cal Y}_L}}
\end{bmatrix}} {\rm{,}}\\
{{\cal M}_\kappa }\left( t \right) = {\rm{det}}\left\{ {{\rm{bdiag}}\left\{ {{{\cal M}_{AB}}{I_{{n_\Theta }}}{\rm{,\;}}{{\cal M}_L}{I_{n_{e}}}} \right\}} \right\}{\rm{,}}
\end{gather*}
where the following proposition holds for ${{\cal M}_\kappa }\left( t \right)$.

{\bf{Proposition 1.}} \emph{If $\overline \varphi \left( t \right) \in {\rm{FE}}$ over the time range $\left[ {{t_{\epsilon}}{\rm{;\;}}{t_e}} \right]$, then for all $t \ge {t_e}$ it holds that ${{\cal M}_\kappa }\left( t \right) \ge \underline {{{\cal M}_\kappa }}  > 0$.}

\emph{Proof of Proposition 1 is given in Appendix.}

Thus, under the condition that Hypotheses 1 and 2 are met, we have obtained the LRE \eqref{eq23} without application of the division operation. Such LRE has regressand ${{\cal Y}_\kappa }\left( t \right)$ and regressor ${{\cal M}_\kappa }\left( t \right)$, which are measurable on the basis of signals $u\left( t \right)$ and $y\left( t \right)$. Now we are in position to derive procedure to obtain the estimates ${\hat \Theta _{AB}}\left( t \right)$ and $\hat L\left( t \right)$.

\subsection{Identification and Observation}

In order to analyze the properties of the observation error ${\tilde x_e}\left( t \right)$, the error equation between \eqref{eq7} and \eqref{eq5} is written as:
\begin{equation}\label{eq24}
\begin{array}{l}
{{\dot {\tilde{ x}}}_e}\left( t \right) =  - {\Phi ^{\rm{T}}}\left( {{x_e}{\rm{, }}u} \right){\Theta _{AB}}\left( \theta  \right) - {A_\delta }{x_e}\left( t \right) + \\
 + {\Phi ^{\rm{T}}}\left( {{{\hat x}_e}{\rm{, }}u} \right){{\hat \Theta }_{AB}}\left( t \right) + {A_\delta }{{\hat x}_e}\left( t \right) - \hat L\left( t \right)\tilde y\left( t \right) = \\
 = {\Phi ^{\rm{T}}}\left( {{{\hat x}_e}{\rm{, }}u} \right){{\hat \Theta }_{AB}}\left( t \right) + {A_\delta }{{\tilde x}_e}\left( t \right) - \hat L\left( t \right)\tilde y\left( t \right) - \\
 - {\Phi ^{\rm{T}}}\left( {{x_e}{\rm{, }}u} \right){\Theta _{AB}}\left( \theta  \right) \pm {\Phi ^{\rm{T}}}\left( {{{\hat x}_e}{\rm{, }}u} \right){\Theta _{AB}}\left( \theta  \right) = \\
 = \left( {{A_e}\left( \theta  \right) + {A_\delta }} \right){{\tilde x}_e}\left( t \right) + {\Phi ^{\rm{T}}}\left( {{{\hat x}_e}{\rm{, }}u} \right){{\tilde \Theta }_{AB}}\left( t \right) - \\
 - \hat L\left( t \right)\tilde y\left( t \right) \pm L\tilde y\left( t \right) = \\
 = {A_m}{{\tilde x}_e}\left( t \right) + {\Phi ^{\rm{T}}}\left( {{{\hat x}_e}{\rm{, }}u} \right){{\tilde \Theta }_{AB}}\left( t \right) - \tilde L\left( t \right)\tilde y\left( t \right) = \\
{\rm{ = }}{A_m}{{\tilde x}_e}\left( t \right) + {\phi ^{\rm{T}}}\left( t \right)\tilde \kappa \left( t \right){\rm{,}}
\end{array}
\end{equation}
where
\begin{gather*}
    {\phi ^{\rm{T}}}\left( t \right) = {\begin{bmatrix}
{{\Phi ^{\rm{T}}}\left( {{{\hat x}_e}{\rm{, }}u} \right)}&{ - \tilde y\left( t \right){I_{n_{e}}}}
\end{bmatrix}}{\rm{,}}
\end{gather*}
and $\tilde y\left( t \right) = \hat y\left( t \right) - y\left( t \right)$, ${A_m} = {A_e}\left( \theta  \right) + {A_\delta } - L\left( \theta  \right)C_e^{\rm{T}}$ is a Hurwitz matrix according to Assumption 3.

Having at hand measurable LRE \eqref{eq23}, which scalar regressor ${{\cal M}_\kappa }\left( t \right)$ is bounded away from zero for all $t \ge {t_e}$, an estimation law is derived on the basis of the results from \cite{b11} to ensure exponential stability of system \eqref{eq24} and guarantee that the goal \eqref{eq8} is achieved.

{\bf{Theorem 1.}} \emph{If $\overline \varphi \left( t \right) \in {\rm{FE}}$ over the time range $\left[ {{t_{\epsilon}}{\rm{;\;}}{t_e}} \right]$ and ${\gamma _0} > 0$, ${\gamma _1} > 0$ , then the estimation law:}
\begin{equation}\label{eq25}
\begin{array}{c}
\dot {\hat {\kappa}} \left( t \right) =  - \gamma \left( t \right){{\cal M}_\kappa }\left( t \right)\left( {{{\cal M}_\kappa }\left( t \right)\hat \kappa \left( t \right) - {{\cal Y}_\kappa }\left( t \right)} \right) = \\
 =  - \gamma \left( t \right){\cal M}_\kappa ^{\rm{2}}\left( t \right)\tilde \kappa \left( t \right){\rm{, }}\\
\gamma \left( t \right){\rm{:}} = \left\{ \begin{array}{l}
0,{\rm{\;if\;}}\Delta \left( t \right) < \rho  \in \left[ {{\Delta _{\min}}{\rm{;\;}}{\Delta _{\max }}} \right){\rm{,}}\\
\frac{{{\gamma _0}{\lambda _{{\rm{max}}}}\left( {\phi \left( t \right){\phi ^{\rm{T}}}\left( t \right)} \right) + {\gamma _1}}}{{{\cal M}_\kappa ^2\left( t \right)}}{\rm{\;otherwise}}{\rm{,}}
\end{array} \right.
\end{array}
\end{equation}
\emph{ensures exponential convergence of ${{\begin{bmatrix}
{{{\tilde x}^{\rm{T}}}\left( t \right)}&{{{\tilde \kappa }^{\rm{T}}}\left( t \right)}
\end{bmatrix}}^{\rm{T}}}$ to zero for all $t \ge {t_e}$.}

\emph{Proof of Theorem 1 is verbatim to the one of Theorem from \cite{b11}.}

Therefore, implementation of the law \eqref{eq25} to adjust the observer \eqref{eq7} parameters is based on: 1) filtering \eqref{eq12}-\eqref{eq14} augmented with a regressor mixing procedure to obtain the LRE \eqref{eq11} with respect to the parameters $\eta \left( \theta  \right)$ with a scalar regressor $\Delta \left( t \right)$; 2) mappings \eqref{eq18}, \eqref{eq20}, \eqref{eq21} to transform the LRE with respect to the parameters $\eta \left( \theta  \right)$ to the one with respect to $\kappa$.

\section{Numerical Experiments}

We have considered the system from the experimental section of \cite{b11} and augmented it with an external disturbance:
\begin{equation}\label{eq26}
\begin{gathered}
  \dot x = {\begin{bmatrix}
  0&{{\theta _1} + {\theta _2}}&0 \\ 
  { - {\theta _2}}&0&{{\theta _2}} \\ 
  0&{ - {\theta _3}}&0 
\end{bmatrix}} x + {\begin{bmatrix}
  0 \\ 
  0 \\ 
  {{\theta _3}} 
\end{bmatrix}} u + {\begin{bmatrix}
  {{\theta _1}{\theta _2}} \\ 
  0 \\ 
  0 
\end{bmatrix}}\delta  = \\
= {\begin{bmatrix}
  {{I_3}}&{{0_{3 \times 2}}} 
\end{bmatrix}} {\Phi ^{\text{T}}}\left( {{x_e}{\text{, }}u} \right) {\begin{bmatrix}
  {{\theta _1} + {\theta _2}} \\ 
  {{\theta _1}{\theta _2}} \\ 
  {{\theta _2}} \\ 
  {{\theta _3}} 
\end{bmatrix}} {\text{,}} \hfill \\
  y = {\begin{bmatrix}
  0&0&1 
\end{bmatrix}} x{\text{,}} \hfill \\ 
\end{gathered}    
\end{equation}
where 
\begin{displaymath}
{\begin{bmatrix}
  {{I_3}}&{{0_{3 \times 2}}} 
\end{bmatrix}}{\Phi ^{\text{T}}}\left( {{x_e}{\text{, }}u} \right) =  {\begin{bmatrix}
  {{x_{2e}}}&{{x_{4e}}}&0&0 \\ 
  0&0&{{x_{3e}} - {x_{1e}}}&0 \\ 
  0&0&0&{u - {x_{2e}}} 
\end{bmatrix}}.
\end{displaymath}

The equation \eqref{eq26} was transformed into the observer canonical form \eqref{eq9} with the following parameter vectors:
\begin{equation}\label{eq27}
\begin{gathered}
{\psi _a} = {\begin{bmatrix}
  0 \\ 
  { - \left( {{\theta _1} + {\theta _2} + {\theta _3}} \right){\theta _2}} \\ 
  0 
\end{bmatrix}} {\text{, }}\\
{\psi _b} = {\begin{bmatrix}
  {{\theta _3}} \\ 
  0 \\ 
  {{\theta _3}{\theta _2}\left( {{\theta _2} + {\theta _1}} \right)} 
\end{bmatrix}}{\text{, }}{\psi _d} = {\begin{bmatrix}
  0 \\ 
  0 \\ 
  {{\theta _1}\theta _2^2{\theta _3}} 
\end{bmatrix}},
\end{gathered}
\end{equation}
where ${\psi _{ab}}\left( \theta  \right) = col\left\{ { - \left( {{\theta _1} + {\theta _2} + {\theta _3}} \right){\theta _2}{\text{, }}{\theta _3}{\text{, }}{\theta _3}{\theta _2}\left( {{\theta _2} + {\theta _1}} \right)} \right\}.$

The parameters of the exosystem \eqref{eq4} were set as:
\begin{equation}\label{eq28}
\begin{gathered}
{\mathcal{A}_\delta } = {\begin{bmatrix}
  0&1 \\ 
  { - 10}&{ - 0.{\text{01}}} 
\end{bmatrix}} {\text{, }}h_\delta ^{\text{T}} = {\begin{bmatrix}
  1&0 
\end{bmatrix}}.
\end{gathered}
\end{equation}

The mappings ${\mathcal{T}_\mathcal{S}}\left( . \right){\text{, }}{\mathcal{T}_\mathcal{G}}\left( . \right){\text{, }}{\mathcal{T}_\Theta }\left( . \right)$ were implemented as follows:
\begin{displaymath}
\begin{gathered}
  {\mathcal{T}_\mathcal{S}}\left( {{{\overline \Xi }_\mathcal{S}}\left( \Delta  \right){\mathcal{Y}_{ab}}} \right) \!=\! {\begin{bmatrix}
  {{\mathcal{Y}_{2ab}}{{\left( {{\mathcal{Y}_{1ab}}{\mathcal{Y}_{2ab}} \!+\! \Delta {\mathcal{Y}_{3ab}}} \right)}^2} \!-\! \mathcal{Y}_{2ab}^4{\mathcal{Y}_{3ab}}} \\ 
  { - {\mathcal{Y}_{1ab}}{\mathcal{Y}_{2ab}} - \Delta {\mathcal{Y}_{3ab}}} \\ 
  {{\mathcal{Y}_{2ab}}{\mathcal{Y}_{1ab}}} 
\end{bmatrix}}{\text{,}}\\
{\mathcal{T}_\mathcal{G}}\left( {{{\overline \Xi }_\mathcal{G}}\left( \Delta  \right){\mathcal{Y}_{ab}}} \right) = {\begin{bmatrix}
  {\mathcal{Y}_{2ab}^3\left( {{\mathcal{Y}_{1ab}}{\mathcal{Y}_{2ab}} + \Delta {\mathcal{Y}_{3ab}}} \right)} \\ 
  {\mathcal{Y}_{2ab}^2} \\ 
  {\Delta {\mathcal{Y}_{1ab}}} 
\end{bmatrix}}{\text{,  }} \\ 
  {\mathcal{T}_\Theta }\left( {{\Xi _\Theta }\left( \Delta  \right){\mathcal{Y}_\theta }} \right) = {{\begin{bmatrix}
  {{\mathcal{Y}_{1\theta }} + {\mathcal{Y}_{2\theta }}}&{{\mathcal{Y}_{1\theta }}{\mathcal{Y}_{2\theta }}}&{{\mathcal{Y}_{2\theta }}}&{{\mathcal{Y}_{3\theta }}}
\end{bmatrix}}^{\text{T}}}.
\end{gathered} 
\end{displaymath}

The values of the functions ${\mathcal{Y}_L}\left( t \right){\text{, }}{\mathcal{M}_L}\left( t \right)$ were calculated by \eqref{eq21} using ${\mathcal{Y}_{AB}}\left( t \right)$. The initial conditions of the extended system \eqref{eq26}, \eqref{eq4}, parameters of filters \eqref{eq12}-\eqref{eq14}, exosystem \eqref{eq6} and estimation law \eqref{eq25} were set as:
\begin{displaymath}
\begin{gathered}
  K = {{\begin{bmatrix}
  3&3&1 
\end{bmatrix}} ^{\text{T}}}{\text{, }}G = {\begin{bmatrix}
  { - 4}&1 \\ 
  { - 2}&0 
\end{bmatrix}}{\text{, }}l = {\begin{bmatrix}
  1 \\ 
  2 
\end{bmatrix}}{\text{, }}\beta  = {\begin{bmatrix}
  {20} \\ 
  { - 8} 
\end{bmatrix}}{\text{,}} \\ 
  {x_{0e}} = {{\begin{bmatrix}
  { - 1}&0&2&{500}&{100} 
\end{bmatrix}} ^{\text{T}}}{\text{, }}\\
\kappa  = {{\text{0}}_9}{\text{, }}\sigma \left( \Gamma  \right) =  - {1_{n_{e}}}{\text{, }} k = {10^{19}}{\text{, }}{k_1} = 25{\text{, }}k_{2}  = 1{\text{, }}\\
{t_{\epsilon}} = 25{\text{, }}\rho  = 0{\text{.1, }}{\gamma _1} = 1{\text{, }}{\gamma _0} = {10^{ - 11}}. \\ 
\end{gathered}
\end{displaymath}

The control law was formed as P-controller $u \!=\!  - 75\left( {r \!-\! y} \right)$ with the reference $r = 100 + 2.5{e^{ - \int\limits_{{t_\epsilon}}^t {d\tau } }}\sin \left( {10t} \right) $. For comparison purposes the non-adaptive version of the observer \eqref{eq7} was implemented:
\begin{equation}\label{eq29}
\begin{gathered}
\dot {\hat x}_e^ * \left( t \right) = {\Phi ^{\text{T}}}\left( {\hat x_e^ * {\text{, }}u} \right){\Theta _{AB}} + {A_\delta }\hat x_e^ *  - L\left( {C_0^{\text{T}}\hat x_e^ *  - y} \right).
\end{gathered}
\end{equation}
	
Figure 1 depicts $\overline \varphi _f^{\rm{T}}\left( t \right)\eta \left( \theta  \right)$ and ${\overline \varepsilon _f}\left( t \right) + {e^{ - {k_1}\left( {t - {t_0}} \right)}}\overline q\left( {{t_0}} \right)$.

\begin{figure}[htbp]
\centerline{\includegraphics[scale=0.58]{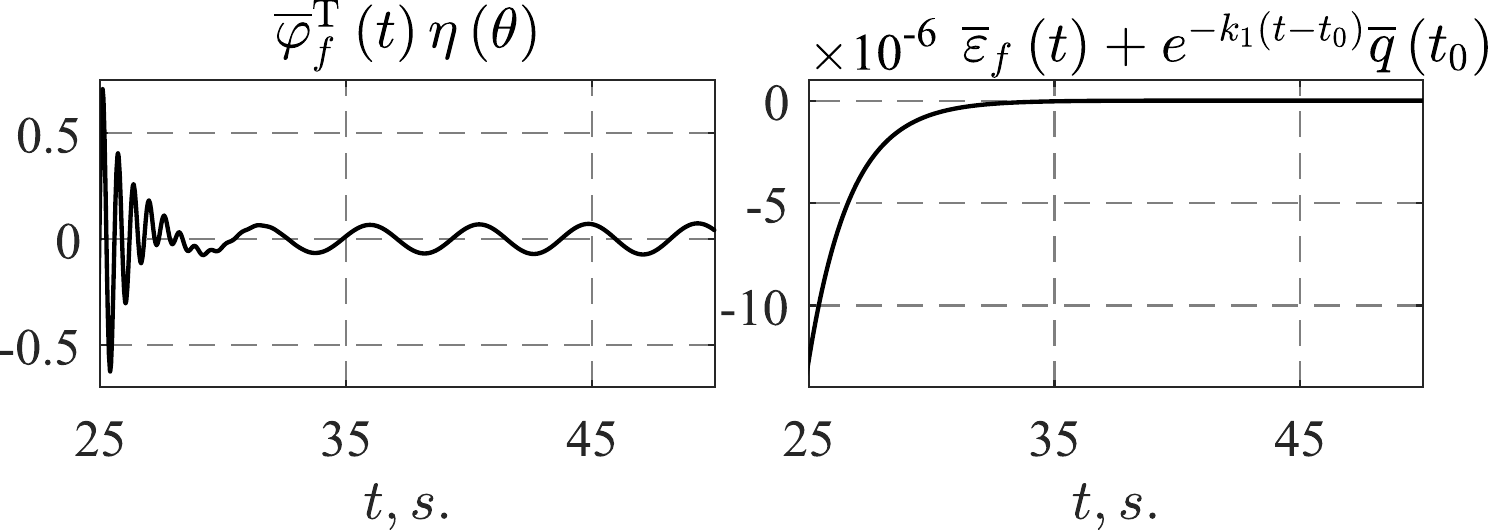}}
	\caption{Comparison $\overline \varphi _f^{\rm{T}}\left( t \right)\eta \left( \theta  \right)$ and ${\overline \varepsilon _f}\left( t \right) + {e^{ - {k_1}\left( {t - {t_0}} \right)}}\overline q\left( {{t_0}} \right)$}
	\label{fig1}
\end{figure}

Transient curves in Figure 1 validate that the condition $\varepsilon \left( t \right) = o\left( {\varphi \left( t \right)\eta(\theta) } \right)$ from Lemma 1 was met for all $t \geqslant {t_e} = 25{\text{ s}}{\text{.}}$ Figure 2 presents transients of ${\tilde \Theta _{AB}}$ and normalized errors ${{{{\left( {{\hat{x}_i} - \hat x_i^ * } \right)}}} \mathord{\left/
 {\vphantom {{{{\left( {{\hat{x}_i} - \hat x_i^ * } \right)}_i}} {\mathop {\max }\limits_{t \in \left[ {0{\text{; 50}}} \right]} \left| {{\hat{x}_i} - \hat x_i^ * } \right|}}} \right.
 \kern-\nulldelimiterspace} {\mathop {\max }\limits_{t \in \left[ {0{\text{; 50}}} \right]} \left| {{\hat{x}_i} - \hat x_i^ * } \right|}}{\text{, }}{{\left( {\hat \delta  - {{\hat \delta }^ * }} \right)} \mathord{\left/
 {\vphantom {{\left( {\hat \delta  - {{\hat \delta }^ * }} \right)} {\mathop {\max }\limits_{t \in \left[ {0{\text{; 75}}} \right]} }}} \right.
 \kern-\nulldelimiterspace} {\mathop {\max }\limits_{t \in \left[ {0{\text{; 50}}} \right]} }}\left| {\hat \delta  - {{\hat \delta }^ * }} \right|$ and ${\text{ }}{{{{\tilde L}_i}} \mathord{\left/
 {\vphantom {{{{\tilde L}_i}} {\mathop {\max }\limits_{t \in \left[ {0{\text{; 50}}} \right]} \left| {{{\tilde L}_i}} \right|}}} \right.
 \kern-\nulldelimiterspace} {\mathop {\max }\limits_{t \in \left[ {0{\text{; 50}}} \right]} \left| {{{\tilde L}_i}} \right|}}$.

\begin{figure}[htbp]
\centerline{\includegraphics[scale=0.58]{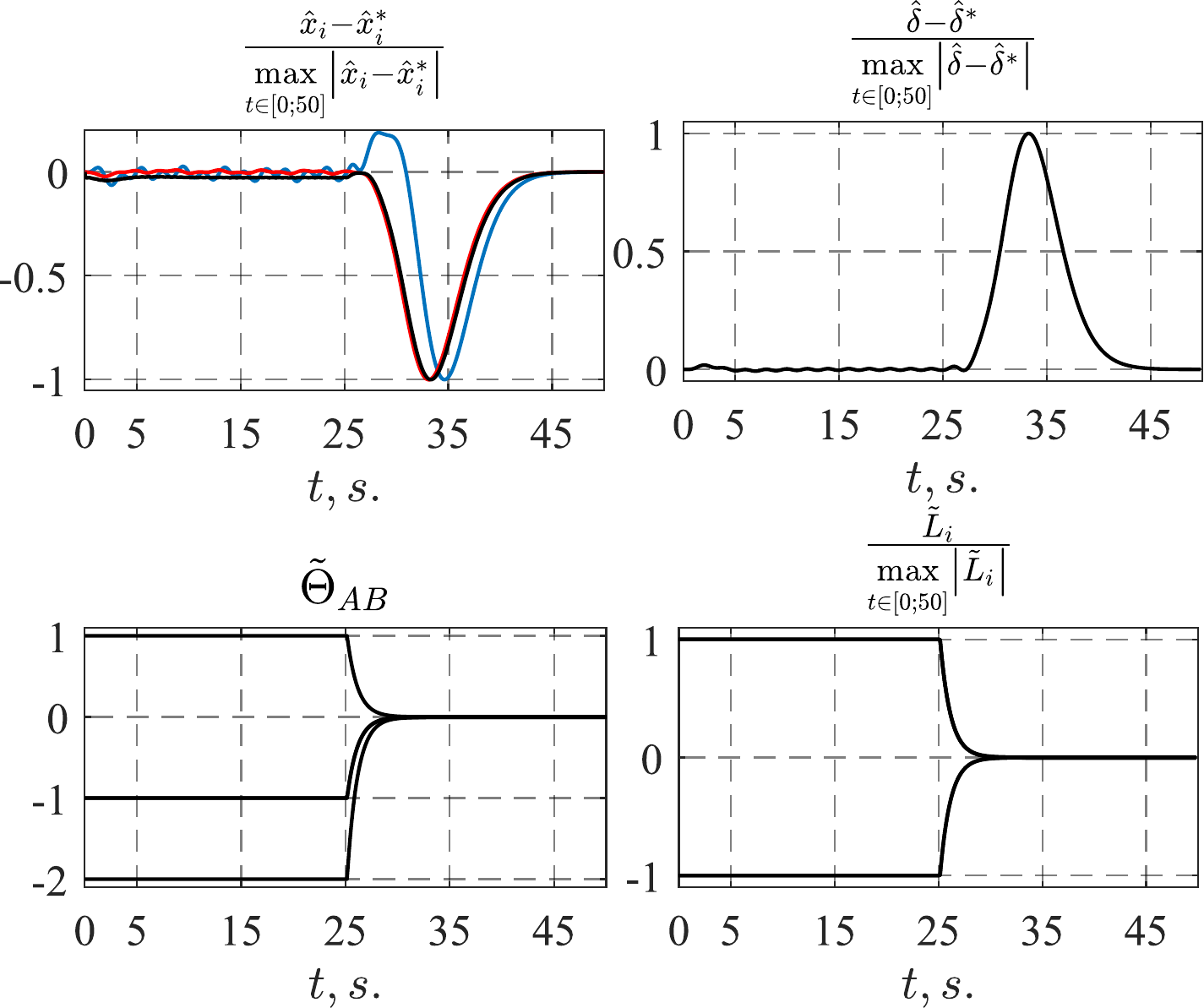}}
	\caption{Transients of estimation and identification errors}
	\label{fig2}
\end{figure}

Therefore, when the parameter estimation process was completed, the estimates ${\hat x_e}\left( t \right)$ formed by the adaptive observer \eqref{eq7} converged to the ones obtained with the help of the observer \eqref{eq29} with ideal parameters. Considering ${\hat x_e}\left( t \right)$ transients, the overshoot is explained by the peaking phenomenon \cite{b14}, which occurs when a linear system with nonzero initial conditions is included into a feedback loop with high gains. In general, the results of the experiment demonstrated that the goal \eqref{eq8} was achieved.

\section{Conclusion}

An extended adaptive observer is proposed for a class of linear systems with overparameterization, which, if the regressor is finitely exciting, allows one to reconstruct unmeasured state and bounded external perturbation produced by a known linear exosystem with unknown initial conditions.

In contrast to the solutions from \cite{b1, b2, b3, b4, b5, b6}, the proposed observer ({\it i}) allows one to reconstruct the original state $x\left( t \right)$ of the system \eqref{eq3} rather than the virtual one $\xi \left( t \right)$ of the observer canonical form \eqref{eq9}, ({\it ii}) ensures exponential convergence of the observation error for the extended system state ${x_e}\left( t \right)$ in case the regressor finite excitation requirement is met. Unlike the earlier result \cite{b11}, the above-presented adaptive observer: \linebreak {\it a}) reconstructs the disturbance exosystem \eqref{eq4} states in addition to $x\left( t \right)$, {\it b}) uses a simpler approach to parameterize the LRE with respect to $L\left( \theta  \right)$.

\appendix
\renewcommand{\theequation}{A\arabic{equation}}
\setcounter{equation}{0}  

\subsection{Proof of Lemma 1} 
Following the results of Lemma 1 and Theorem 2 from \cite{b15}, the below-given equation is parametrized with the help of filters \eqref{eq14}:
\begin{equation}\label{eqA1}
\dot {\overline q} = {\overline \varphi ^{\text{T}}}\left( t \right)\eta \left( \theta  \right) + {\beta ^{\text{T}}}\left( {F\left( t \right) + ly\left( t \right)} \right) + \overline \varepsilon \left( t \right),    
\end{equation}
where $\overline \varepsilon \left( t \right)$ denotes aggregated exponentially decaying functions.

The error $\chi \left( t \right) = \overline q\left( t \right) - {k_1}{\overline q_f}\left( t \right)$ is differentiated, and, using \eqref{eqA1} and \eqref{eq13}, it is obtained:
\begin{equation}\label{eqA2}
\begin{gathered}
  \dot {\chi} \left( t \right) = {{\overline \varphi }^{\text{T}}}\left( t \right)\eta \left( \theta  \right) + {\beta ^{\text{T}}}\left( {F\left( t \right) + ly\left( t \right)} \right) + \overline \varepsilon \left( t \right) -\\
  -{k_1}\left( { - {k_1}{{\overline q}_f}\left( t \right) + \overline q\left( t \right)} \right) =  - {k_1}\chi \left( t \right) + {{\overline \varphi }^{\text{T}}}\left( t \right)\eta \left( \theta  \right) +\\
  +{\beta ^{\text{T}}}\left( {F\left( t \right) + ly\left( t \right)} \right) + \overline \varepsilon \left( t \right). 
\end{gathered}    
\end{equation}

Taking into consideration the solution of \eqref{eqA2}, it is written:
\begin{equation}\label{eqA3}
\begin{gathered}
\overline q\left( t \right) - {k_1}{\overline q_f}\left( t \right) - {\beta ^{\text{T}}}\left( {{F_f}\left( t \right) + l{y_f}\left( t \right)} \right) =\\
={e^{ - {k_1}\left( {t - {t_0}} \right)}}\overline q\left( {{t_0}} \right) + \overline \varphi _f^{\text{T}}\left( t \right)\eta\left( \theta  \right)  + {\overline \varepsilon _f}\left( t \right){\text{,}}
\end{gathered}    
\end{equation}
where ${\dot {\overline \varepsilon} _f}\left( t \right) =  - {k_1}{\overline \varepsilon _f}\left( t \right) + {k_1}\overline \varepsilon \left( t \right){\text{, }}{\overline \varepsilon _f}\left( {{t_0}} \right) = 0.$

Owing to \eqref{eqA3}, the solution of the first differential equation from \eqref{eq12} is represented as:
\begin{equation}\label{eqA4}
\begin{gathered}
q\left( t \right) = \varphi \left( t \right)\eta \left( \theta  \right) + \varepsilon \left( t \right){\text{,}}
\end{gathered}    
\end{equation}
where $\dot \varepsilon \left( t \right) = \int\limits_{{t_\epsilon}}^t {{e^{ - k_{2} \tau }}{{\overline \varphi }_f}\left( \tau  \right)\left( {{{\overline \varepsilon }_f}\left( \tau  \right) + {e^{ - {k_1}\left( {\tau  - {t_0}} \right)}}\overline q\left( {{t_0}} \right)} \right)d\tau }$, $\varepsilon \left( {{t_\epsilon}} \right) = {0_{{\text{2}}n}}.$

The disturbance ${\overline \varepsilon _f}\left( \tau  \right) + {e^{ - {k_1}\left( {\tau  - {t_0}} \right)}}\overline q\left( {{t_0}} \right)$ decays exponentially for all $t \geqslant {t_0}$, therefore, if the condition ${t_\epsilon} \gg {t_0}$ is met, then equation \eqref{eqA4} is rewritten as:
\begin{equation}\label{eqA5}
\begin{gathered}
q\left( t \right) = \varphi \left( t \right)\eta \left( \theta  \right) + o\left( {\varphi \left( t \right)\eta \left( \theta  \right)} \right){\text{,}}
\end{gathered}    
\end{equation}
i.e., the contribution of $\varepsilon \left( t \right)$ into $q\left( t \right)$ is small to negligible $q\left( t \right){\text{:}} = \varphi \left( t \right)\eta \left( \theta  \right)$ in case $t_\epsilon$ is chosen to be sufficiently large. Multiplying $q\left( t \right)$ by $k\left( t \right) \cdot {\text{adj}}\left\{ {\varphi \left( t \right)} \right\}$ and applying the property ${\text{adj}}\left\{ {\varphi \left( t \right)} \right\}\varphi \left( t \right) = {\text{det}}\left\{ {\varphi \left( t \right)} \right\}{I_{2n}}$, equation \eqref{eq11} is obtained.

As the signals $y\left( t \right){\text{, }}u\left( t \right)$ are bounded according to Assumption 1, then for all $t \geqslant {t_0}$ the inequality ${\Delta _{{\text{max}}}} \geqslant \Delta \left( t \right)$ holds because the filters \eqref{eq13}, \eqref{eq14} are stable and the integrands from \eqref{eq12} decay exponentially. In accordance with Lemma 6.8 from \cite{b10}, if $\overline \varphi \left( t \right) \in {\text{FE}}$, then it also holds that ${\overline \varphi _f}\left( t \right) \in {\text{FE}}$. Following Theorem 1 from \cite{b16}, if ${\overline \varphi _f}\left( t \right) \in {\text{FE}}$, then for all $t \geqslant {t_e}$ it holds that $\varphi \left( t \right) \!>\! 0 \Leftrightarrow \Delta \left( t \right) \!\geqslant\! {\Delta _{{\text{min}}}} \!>\! 0$, which was to be proved in Lemma 1.

\subsection{Proof of Lemma 2} 
If Assumption 3 is met, in accordance with the results of generalized pole placement theory \cite{b12, b13} the vector $L\left( \theta  \right)$ is obtained using the following set of equations:
\begin{equation}\label{eqA6}
\begin{gathered}
\left\{ \begin{gathered}
  A_e^{\text{T}}\left( \theta  \right)M + A_\delta ^{\text{T}}M - M\Gamma  = {C_e}B_e^{\text{T}}\left( \theta  \right){\text{,}} \hfill \\
  B_e^{\text{T}}\left( \theta  \right) = {L^{\text{T}}}\left( \theta  \right)M{\text{,}} \\ 
\end{gathered}  \right.
\end{gathered}    
\end{equation}
which has unique solution \cite{b12, b13} as, following Assumption 3, the pair $\left( {A_e^{\text{T}}\left( \theta  \right) + A_\delta ^{\text{T}}{\text{, }}{C_e}} \right)$ is controllable, the pair $\left( {B_e^{\text{T}}\left( \theta  \right){\text{, }}\Gamma } \right)$ is observable and $\sigma \left\{ {A_e^{\text{T}}\left( \theta  \right) + A_\delta ^{\text{T}}} \right\} \cap \sigma \left\{ \Gamma  \right\} = 0$.

The properties of the vectorization operation are used to express $M$ from the first equation of \eqref{eqA6}, then the obtained result is substituted into the second equation of \eqref{eqA6}:
\begin{equation}\label{eqA7}
\begin{gathered}
{\cal Q}\left( \vartheta  \right) = {\cal P}\left( \vartheta  \right)L\left( \theta  \right){\rm{,}}\\
\end{gathered}
\end{equation}
\begin{displaymath}
\begin{gathered}
    {\cal Q}\left( \vartheta  \right) = {\rm{det}}\left\{ {{I_{{n_e}}} \otimes \left( {A_e^{\rm{T}}\left( \theta  \right) + A_\delta ^{\rm{T}}} \right) - {\Gamma ^{\rm{T}}} \otimes {I_{{n_e}}}} \right\}{B_e}\left( \theta  \right){\rm{,}}\\
\begin{array}{l}
{\cal P}\left( \vartheta  \right) = ve{c^{ - 1}}\left\{ {{\rm{adj}}\left\{ {{I_{{n_e}}} \otimes \left( {A_e^{\rm{T}}\left( \theta  \right) + A_\delta ^{\rm{T}}} \right) - {\Gamma ^{\rm{T}}} \otimes {I_{{n_e}}}} \right\}} \right.  \\
 \times {\left. {vec\left( {{C_e}B_e^{\rm{T}}\left( \theta  \right)} \right)} \right\}^{\rm{T}}}{\rm{,}}
\end{array}
\end{gathered}
\end{displaymath}

As the following equalities hold for ${\Theta _{AB}}\left( \theta  \right)$ and ${A_e}\left( \theta  \right){\text{, }}{B_e}\left( \theta  \right)$ on the basis of Hypothesis 1:
\begin{displaymath}
\begin{gathered}
{\mathcal{M}_{AB}}A_e^{\text{T}}\left( \theta  \right) = ve{c^{ - 1}}\left( {{\mathcal{L}_{{A^{\text{T}}}}}{\mathcal{D}_\Phi }{\mathcal{Y}_{AB}}} \right){\text{, }}\\
{\mathcal{M}_{AB}}{B_e}\left( \theta  \right) = {\mathcal{L}_B}{\mathcal{D}_\Phi }{\mathcal{Y}_{AB}}{\text{,}} 
\end{gathered}
\end{displaymath}
then multiplication of \eqref{eqA7} by ${\Pi _L}\left( {{\mathcal{M}_{AB}}} \right) = \mathcal{M}_{AB}^{{{\left( {n_{e}} \right)}^2} + 1}{I_{n_{e}}}$ and further application of the properties ${\operatorname{c} ^n}{\text{det}}\left\{ A \right\} = \linebreak={\text{det}}\left\{ {cA} \right\}{\text{, }}{\operatorname{c} ^{n - 1}}{\text{adj}}\left\{ A \right\} = {\text{adj}}\left\{ {cA} \right\}{\text{, }}A \in {\mathbb{R}^{n \times n}}$ allows one to obtain \eqref{eq21}.

\subsection{Proof of Proposition 1}
Following the result of Lemma 1, if $\overline \varphi \left( t \right) \in {\text{FE}}$, then for all $t \geqslant {t_e}$ the inequality $\Delta \left( t \right) \geqslant {\Delta _{{\text{min}}}}$ takes place. At the same time, the below-given inequalities hold according to Hypotheses 1 and 2:
\begin{displaymath}
\begin{gathered}
{\text{de}}{{\text{t}}^2}\left\{ {\mathcal{G}\left( {{\psi _{ab}}} \right)} \right\} > 0{\text{, det}}\left\{ {{\Pi _\theta }\left( {\Delta \left( t \right)} \right)} \right\} \geqslant {\Delta ^{{\ell _\theta }}}\left( t \right){\text{, }} \\{\mathcal{M}_{AB}}\left( t \right) = {\text{det}}\left\{ {{\Pi _\Theta }\left( {{\mathcal{M}_\theta }\left( t \right)} \right)} \right\} \geqslant \mathcal{M}_\theta ^{{\ell _\Theta }}\left( t \right).
\end{gathered}
\end{displaymath}

Owing to solvability of the set of equations \eqref{eqA6}, it is also true that ${\det ^2}\left\{ {\mathcal{P}\left( \vartheta  \right)} \right\} > 0.$ If $\overline \varphi \left( t \right) \in {\text{FE}}$, then for all $t \geqslant {t_e}$ the inequalities hold:
\begin{displaymath}
\begin{gathered}
  \left| {{\mathcal{M}_{AB}}\left( t \right)} \right| = \left| {{\text{det}}\left\{ {{\Pi _\Theta }\left( {{\mathcal{M}_\theta }\left( t \right)} \right)} \right\}} \right| \geqslant \left| {\mathcal{M}_\theta ^{{\ell _\Theta }}\left( t \right)} \right| \geqslant \\
  \geqslant \left| {{\text{de}}{{\text{t}}^{{\ell _\Theta }}}\left\{ {\mathcal{G}\left( {{\psi _{ab}}} \right)} \right\}} \right|\Delta _{\min }^{{\ell _\theta }{\ell _\Theta }} = {\underline {\mathcal{M} _{AB}}} > 0, \\ 
  \left| {{\mathcal{M}_L}\left( t \right)} \right| = \left| {{\text{det}}\left\{ {{\mathcal{T}_\mathcal{P}}\left( {{\Xi _\mathcal{P}}\left( {{\mathcal{M}_\theta }\left( t \right)} \right)\theta } \right)} \right\}} \right| = \\
  =\left| {{\text{det}}\left\{ {\mathcal{P}\left( \theta  \right)} \right\}{\text{det}}\left\{ {\mathcal{M}_{AB}^{{{{n^2_{e}} }} + 1}{I_{n_{e}}}} \right\}} \right| =  \\
  = \left| {{\text{det}}\left\{ {\mathcal{P}\left( \theta  \right)} \right\}} \right|\underline {\mathcal{M} _{AB}}^{\left( {{{{n^2_{e}}}} + 1} \right)\left( {n_{e}} \right)} \geqslant \underline {{\mathcal{M}_L}}  > 0{\text{,}}\\
  \left| {{\mathcal{M}_\kappa }\left( t \right)} \right| = \left| {\mathcal{M}_{AB}^{{n_\Theta }}\left( t \right)\mathcal{M}_L^{n_{e}}\left( t \right)} \right| \geqslant \\
  \geqslant \underline {\mathcal{M}_{AB}^{{n_\Theta }}} \underline {\mathcal{M}_L^{n_{e}}}  = \underline {{\mathcal{M}_\kappa }}  > 0.
\end{gathered}    
\end{displaymath}

\end{document}